# MEASURING THE FORCE EJECTING DNA FROM PHAGE.

Alex Evilevitch, Martin Castelnovo[+], Charles M. Knobler, and William M. Gelbart*

*Department of Chemistry and Biochemistry, University of California, Los Angeles, California 90095-1569, USA*

[+]*Present address: Laboratoire de Dynamique des Fluides Complexes, 3 Rue de l'Universite, Strasbourg, 67000 FRANCE*

ABSTRACT

We discuss how a balance can be established between the force acting to eject DNA from viral capsids and the force resisting its entry into a colloidal suspension which mimics the host cell cytoplasm. The ejection force arises from the energy stored in the capsid as a consequence of the viral genome (double-stranded DNA) being strongly bent and crowded on itself. The resisting force is associated with the osmotic pressure exerted by the colloidal particles in the host solution. Indeed, recent experimental work has demonstrated that the extent of ejection can be progressively limited by increasing the external osmotic pressure; at a sufficiently high pressure the ejection is completely suppressed. We outline here a theoretical analysis that allows a determination of the internal (capsid) pressure by examining the different relations between force and pressure inside and outside the capsid, using the experimentally measured position of the force balance.

1. INTRODUCTION.

Viruses that infect bacteria are among the simplest biological objects[1,2]. Bacterial viruses, or phage, often consist of a single molecule of double-stranded DNA surrounded/protected by a rigid capsid formed from a polyhedral arrangement of many copies of a low-molecular-weight protein. Most phage also have a long, hollow, cylindrical tail that is attached to the hole at a vertex of the capsid. The phage is able to propagate only by getting its DNA inside a cell whose biochemical machinery will do the job of replicating the whole virus – making many copies of its DNA and expressing the genes that are responsible for templating the syntheses of its proteins. After the self assembly of the capsid and tail from their respective constituent proteins, the loading of the DNA into the capsid is directed by a motor protein (also encoded in the viral genes), followed by the subsequent joining of the tail to the phage head. All of these exquisitely orchestrated activities are set into motion upon the initial step of DNA injection into the cytoplasm.

It is not surprising, then, that a great deal of work has been devoted to determining how DNA is transferred from the viral capsid through the tail into the bacterial cell cytoplasm[3-5]. In most cases the process is known to be initiated by the binding of the tail to a "receptor" protein that resides in the outer membrane of the bacterium. This binding leads to a conformational change of the viral tail, opening of the capsid, and subsequent ejection of its contents – the viral DNA – into the cell. The capsid remains outside the cell, as first demonstrated in the classic Hershey-Chase experiment just over 50 years ago.[6] Yet today there is still very little understanding of the mechanical and physical properties of viral capsids and encapsidated DNA, i.e., of the forces and pressures operating while a genome is packaged or ejected. These properties are not only central to elucidating the infection mechanism, but are also relevant to the strength of viral capsids and are of importance for investigating the possibility of using them as containers for non-genomic materials.[7,8]

When a virus is assembled in an infected bacterium, its DNA, which may be tens of micrometers long, must be compressed into a (capsid) volume whose diameter is hundreds of times smaller. More significantly, there is only room to accommodate the DNA inside if the spacing between neighboring chain segments is slightly larger than their diameter. Also, the inner diameter of the capsid is comparable to or often even smaller than the DNA persistence length. Based on these facts it has been recognized that this compressed DNA must exert a very high pressure on the inside of the capsid due to the strong repulsions between the neighboring DNA portions, as well as elastic forces due to the bending of the stiff DNA rod.[9-11] This stored energy density inside the capsid is responsible for the mechanical force that ejects the DNA into the cell, without ATP hydrolysis or the intervention of any other external source of energy.

Recent theoretical work[9-11] has established that the driving force for ejection, $f_{eject}$, decreases monotonically with decreasing length of genome remaining in the capsid. Calculation showed that forces for a fully packaged genome inside the viral capsids reach very high values – tens of pN – corresponding to pressures up to tens of atmospheres exerted by DNA on the capsid walls. The experimental work of Smith et al.[12] is consistent with these predictions. In their single-molecule experiment, the force exerted by the motor protein upon loading of DNA into a phage capsid is measured by stalling the motor with optical tweezers for increasing lengths of packaged DNA. The motor exerts its maximum force at the end of the loading process, when the whole genome is packaged. This force corresponds to the maximum force *ejecting* DNA, at the start of the ejection.

The question asked in recently reported experimental work[13] was whether it is possible to progressively suppress the ejection of the genome from a capsid that is exposed to an increasing osmotic pressure difference between its inside and outside. This situation is present in the *in vivo* situation where the viral genome is ejected into a bacterial cytoplasm whose macromolecular-induced osmotic pressure is several atmospheres.[14-16] In the experiment we mimicked the bacterial cytoplasm with a host solution of an osmotic stress polymer and showed that the osmotic pressure difference, due to the PEG (polyethylene glycol) polymer that remains outside the capsid, gives rise to a resisting force, $f_{resist}$, that balances the force driving ejection (in this case, from the bacteriophage lambda); see the illustration in Fig. 1. Progressively smaller extents of ejection are measured upon increase in the concentration of polymer, ultimately leading to complete inhibition of the ejection. Conversely, in the absence of PEG the ejection is complete because there is no force acting to resist the ejection; this has indeed been demonstrated in many experiments.[17-20] We showed further[13] that the external osmotic pressures necessary to inhibit ejection can be drastically lowered by addition of polyvalent cations, in agreement with many theories (see, example, the references cited in several recent papers[21-24]) and measurements[25-28] on interhelical interactions in DNA condensates and hexagonal phases. We report elsewhere the dependence of critical osmotic pressures on the genome lengths in different phage strains.[29]

Strictly speaking, our osmotic pressure inhibition experiments do not lead to direct measurements of the forces or pressures operative in viral capsids. Rather, they establish the existence of a *balance* between the force acting to eject the DNA from the capsid and the force resisting its entry into the surrounding solution. Furthermore, they show how the *position* of this force balance is controlled by the concentration of osmolyte in solution, e.g., how increasing the PEG concentration leads to the forces becoming balanced with a longer length of genome remaining inside, and hence a larger value of the ejection force. To deduce actual values for the forces, however, requires a systematic theory of the relationship between $f_{eject}$ and the length of DNA remaining inside the capsid, and between $f_{resist}$ and the concentration of osmolyte outside.

Recently we formulated a theory for the force resisting chain insertion into a colloidal suspension, which relates the resisting osmotic force to the concentration of the solute.[30] Our theoretically predicted osmotic force estimates were shown to be of the same order of magnitude as the forces associated with the DNA ejection, as estimated by various calculations of the state of stress of DNA confined in the capsid as a function of genome length inside.[9-11] In the present work we apply these theories and estimates to the case of phage lambda and relate the measured length of ejected genome to the force resisting the ejection; in this way we deduce the force acting to eject the genome as a function of its length remaining inside the capsid. A central goal of this work is to treat the balance between ejecting and resisting forces, thereby explaining the physical meaning and the origin of these forces in relation to the pressures inside and outside the viral capsid. In particular, we account for how the force balance is consistent with a significant difference between inside and outside pressures.

2. FORCE / PRESSURE RELATIONSHIPS.
**A. Force-balance experiment**

The origin of the force resisting the ejection of DNA from the capsid lies in the osmotic pressure caused by molecules in solution that cannot penetrate the capsid. The physics of this osmotic force that resists the entry of DNA into the external solution is the following. In order for the DNA to be inserted into the colloidal suspension, particles have to be removed from an effective volume associated with DNA in the solution. In other words, a cavity must be made in the solution to accommodate entry of this stiff chain. Creating this cavity involves a double energetic price: work must be performed against the pressure of the polymer solution (pressure-volume contribution), and an interface must also be established (surface-tension term). Additionally, formation of the cavity costs some conformational entropy associated with the polymers. The total energy per unit length associated with the insertion of the DNA in the polymer solution gives the *quasistatic force* resisting the ejection, which to leading order does not depend on the length of the DNA chain being inserted into solution.[30] Note that the actual process of DNA ejection is also a dynamical one, and additional effects need to be considered to fully address this problem, e.g., the molecular friction of the DNA inside the phage tail and in the polymer solution. This effective viscosity determines the *rate* of ejection, as the genome is pushed out of the capsid through the tail. *However, all these effects are irrelevant when an equilibrium state is reached*; our experiments demonstrate that there indeed exists a static force capable of balancing the ejection force, resulting in PEG-dependent *partial* ejection of DNA from viral capsids.

The experiment was carried out in the following way[13]. The spontaneous DNA ejection from phage lambda (EMBL3 with a 13kb insert) is triggered *in vitro* by mixing the phage with its receptor protein LamB isolated from the bacterial membrane of *Escherichia Coli*. The DNA is ejected into a buffered aqueous solution of PEG polymer at varying PEG concentrations. The endonuclease DNaseI is also present in the host solution, and it digests all of the DNA that is ejected. This DNA is separated by centrifugation from the phage capsids (with *un*ejected DNA inside), and the concentration of DNA digested into non-sedimenting nucleotides is measured directly by UV absorption at 260 nm. Since the relationship between the osmotic pressure and temperature for PEG8000 solutions for a wide range of concentrations has been well established[25a], the fraction of DNA ejected from the phage can be deduced as a function of the osmotic pressure in the surrounding solution, $\Pi_{out}$; see Fig. 2a. The solid curve in Fig. 2a shows that DNA ejection is monotonically suppressed with increasing osmotic pressure and is completely inhibited at 20 atmospheres, corresponding to 29% wt/wt PEG8000. The time for the ejected DNA to be digested by DNaseI is an order of magnitude longer than the time for its ejection from the phage (minutes[31], versus seconds[19] for the later process). However, the fraction of ejected DNA is measured at long times after complete digestion into nucleotides has occurred and no more DNA is entering the solution from the phage; indeed, no time dependence is observed once the ejected DNA is digested. Therefore, we are establishing a force balance, at equilibrium, between the ejecting force, $f_{eject}$, pushing the DNA from inside the capsid, and the resisting force, $f_{resist}$, exerted by PEG from the outside. We emphasize again that this experiment provides little information about the *non*equilibrium aspects of the *dynamical* process of DNA ejection; nor does its interpretation depend on any of those features. Rather, we show here how $f_{eject}$ can be determined from $f_{resist}$ by analyzing the *equilibrium* nature of an osmotic force balance. To learn the actual values of $f_{eject}$ we need to know the relationships between force and pressure inside and outside the viral capsid. We start with an analysis of the force and pressure resisting the DNA ejection outside the capsid.

**B. Force vs pressure outside the capsid**

The relation between force and pressure in the external solution has been determined in a previous paper.[30] The work needed to create a cavity to accommodate a colloidal particle (e.g., the inserted DNA) in the host solution is commonly decomposed[32] into a sum of a pressure-volume energy, a surface term, and a contribution from the conformational entropy of the osmotic stress polymer. More explicitly, the work of insertion of a rodlike particle of volume $V = \pi(D/2)^2 L$ and surface $S = \pi DL$ (where $L$ is the rod length) is given by

$$W = \Pi_{out}V + \sigma S + cLD^{1/3}b^{5/3}. \qquad (1)$$

Here $\Pi_{out}$, $\sigma$ and $c$ are, respectively, the external osmotic pressure, the surface tension and the monomeric concentration of the polymer solution[32]; $b$ is the monomer size. The last term is important in the DNA case only if its diameter, $D$, is small compared to the characteristic length scale of the polymer solution – $R_g$, the radius of gyration, or $\xi$ the correlation length, in dilute and semi-dilute solution, respectively. In our situation, where $D$ is comparable to $R_g$, the pressure-volume work term is dominant.

Since the calculation of the resisting force uses a scaling description of the polymer solution, it implicitly misses numerical prefactors. However similar calculations used to describe experimental data on the DNA condensation in a PEG solution show that those numerical prefactors can be approximated by unity to within a factor of two.[33] Furthermore, as remarked above, the osmotic force (work of insertion per unit length) is strongly dominated by the leading term involving pressure-volume work. Combining theoretically calculated curves of force vs volume fraction of PEG (see Eq. 23 in ref. 30) with similarly calculated curves of osmotic pressure $\Pi_{out}$ vs volume fraction (see Eqs. 19-22[30]), we obtain the resisting force, $f_{resist}$, as a function of osmotic pressure of the polymer, $\Pi_{out}$. This force, computed with unit prefactors, is shown in Fig. 2b as a function of osmotic pressure for $b=0.4nm$ (approximate size of ethylene glycol monomer)[34] and $M=130$ (degree of polymerization of PEG8000).[34] We use theoretical values of both force and osmotic pressure here in order to be self-consistent in expressing the relation between them. Moreover, since the force and pressure are theoretically determined only up to unknown numerical constants, this procedure minimizes the effect of these scaling prefactors.

The theoretical result shown in Fig. 2b is then combined with the experimental result in Fig. 2a to obtain the ejection force of DNA as a function of genome fraction ejected; see solid curve in Fig. 2c. This force is an increasing function of the fraction of genome inside, and is in qualitative agreement with single-molecule experiments on bacteriophage $\phi29$[12] and with theoretical predictions.[9-11] As is evident from Fig. 2c the ejection force for the fully packaged phage is 18±4 pN (where the uncertainty is that associated with the measured curve[13] shown in 2a). Here it is important to note that the experiment was carried out with a lambda mutant whose genome is 14% shorter than in the wild-type lambda (41.5 kb vs 48.5 kb). Therefore, we would expect to find much higher ejection forces for a fully packaged wild-type phage, which would consequently require a higher osmotic pressure to completely suppress ejection. Experiments and theory illustrating the effect of genome length on ejection forces and pressures will be discussed elsewhere.[29]

We have also performed experiments in which a concentration of 1 mM spermine tetrachloride (+4 cation) was added to the incubating samples of phage, receptor, DNase I and PEG. Polyvalent counterions such as spermine (+4) or spermidine (+3), which are present in significant concentrations in most bacterial cells, act as condensing agents by introducing attractive interaction between the DNA strands[13,25-28]. They reduce the electrostatic repulsion and therefore should decrease the ejection force; the phage capsids are permeable to spermine, just as they are to simple salts (and of course to water). We chose a concentration of 1 mM spermine because this is the minimal amount required to condense free lambda DNA at a concentration corresponding to that of the phage-encapsidated DNA present in our solutions. We find that ejection in the presence of 1 mM spermine is completely suppressed at a pressure of 4 atmospheres (15% w/w PEG8000), much lower than the 20 atmospheres (29% w/w PEG8000) required when no spermine is present; see dot-dash vs solid curves in Fig. 2a. In the presence of spermine a dramatic suppression of the DNA ejection is already evident at only 1 atmosphere osmotic pressure (7% w/w PEG8000), where almost 50% of the genome still remains in the capsid. In the absence of spermine only 15% of the genome remains in the capsid after opening the phage under comparable osmotic pressure. The experimental result presented in Fig. 2a demonstrates that in the presence of spermine a significantly smaller resisting force is sufficient to stop the ejection.

Combining the dot-dash data in Fig. 2a with the calculated force-pressure curve in Fig. 2b allows us, once again, to plot ejection force as a function of fraction of the ejected DNA, now in the presence of added spermine; see dot-dash curve in Fig. 2c. It is evident that the DNA ejection force for a fully packaged genome is now about 5 pN, almost 4 times smaller than in the absence of polyvalent counterions.

## C. Force vs pressure inside the capsid

The numerical estimates of the *resisting force* above can be compared to recent analytical and computational work on the *ejection force* in λ phage.[9-11] In all these cases, the internal force associated with the packaging of DNA inside a spherical viral capsid is computed through the balance of bending and "crowding" energies of DNA. The former energy gives a non-negligible contribution since the size of the capsid is comparable to the persistence length of DNA (about 50nm). But the crowding energy gives by far the largest contribution when the full viral genome is packaged, because of the dominance of short-ranged electrostatic and hydration interactions among neighboring portions of the strongly confined DNA. Of course the interaxial spacing characterizing the packaged genome varies with its (mode of organization and) length. To obtain this relation we use diffraction results on the interaxial spacing as a function of the genome length in fully packaged capsids of different mutants of lambda phage.[35,36] These data suggest a local hexagonal packing and an interaxial spacing of 2.49 nm for the fully packaged EMBL3 mutant. As discussed in previous work,[10] where the bending energy contribution is shown to be relatively small, the ejection force, $f_{eject}(d)$, can be approximated by

$$f_{eject}(d) = \sqrt{3}\int_{d}^{+\infty} dx \Pi_{in}(x) x \qquad (2)$$

Here $\Pi_{in}(d)$ is the osmotic pressure as a function of interaxial distance, $d$, in the hexagonal phase of DNA measured by Rau and Parsegian[25b]. That is, we assume that the osmotic pressure measured for hexagonally ordered DNA in concentrated solutions applies locally to the packaged DNA in phage capsids. In Fig. 3 $f_{eject}(d)$ calculated in this way is plotted for two different sets of solution conditions. We can do this because the viral capsid is fully permeable – not only to water, but also – to all of the salt species involved.

Since the measured pressure equation of state data are not available for precisely the conditions of our buffer solution, namely 10mM $MgSO_4$/50mM TrisCl, we use the $\Pi_{in}(d)$ measured[25b] for the nearest composition – 0.5M NaCl/10mM TrisCl/1mM EDTA. Integrating this $\Pi_{in}(d)$ according to Eq. (2) gives the dashed curve in Fig. 3. The ejection force for DNA inside the fully loaded capsid, i.e., at an inter-axial spacing of 2.49 nm (marked with the vertical dotted line), is about 11 pN (upper horizontal dotted line). Since the ionic strength of the buffer solution used in our experiments was lower than for the result shown in Fig. 3, we expect our measured estimate of the resisting force (see Fig. 2c) to be larger than the value of the ejecting force obtained from Eq. (2) and the Fig. 3 data because in our case there is less screening of the electrostatic repulsion between DNA strands. Consistent with this, our estimate for the resisting force, obtained by using the measured value of the minimum PEG concentration required to completely stop ejection (fully packaged capsid), is 18± 4 (vs 11) pN.

On the other hand, only 4 atm of osmotic pressure is required to completely stop ejection when 1mM tetravalent cation is present in the solution. As in the case where no spermine is present, it is possible to obtain an estimate of the ejection force from the equation of state of hexagonal-phase DNA in the presence of polyvalent ions. The closest available pressure data are those measured for 20 mM $Co(NH_3)_6^{3+}$ with buffer conditions of 0.25 NaCl/10 mM TrisCl[25b]. Integrating this $\Pi_{in}(d)$ as in Eq. (2) we find the dot-dashed curve in Fig. 3. The ejection force is about 2 pN, as shown by the lower horizontal dotted line in Fig. 3, which is once again of the same order of magnitude as our experimentally deduced value (5 pN) for the force resisting ejection in Fig. 2c. Because of the lower ionic strength in our experiment (1mM spermine (+4)/10mM $MgSO_4$/50mM TrisCl) as compared with that used by Parsegian and Rau[25b], we obtain a slightly higher ejection force (almost 5 pN) for the fully packaged capsid, due to the stronger electrostatic repulsion between the DNA strands in the capsid.

For the cases of both simple salt and of polyvalent salt, the estimation of the ejection force (using a combination of the experimental pressure required to completely stop DNA ejection and the theoretical prediction of the associated resisting force) is in qualitative agreement with the forces obtained from integrating the independently measured pressure equations of state. In addition to the ejection force, the theoretical treatments of packaged DNA[9-11] also predict directly the *pressures* associated with this force *inside* the phage; these turn out to be significantly higher than the pressures *outside* that are found sufficient to inhibit ejection. To explain this apparent discrepancy we need to analyze the force versus pressure relationship inside the capsid in comparison with the force-pressure relationship outside the capsid (shown in Fig. 2b).

**D. Inside/outside pressure imbalance**

The pressure that DNA exerts inside the capsid can be calculated by combining the force $f_{eject}(d)$ in Fig. 3 with the pressure measured by Parsegian and Rau[25b] for hexagonally-packed DNA; we identify this pressure with , $\Pi_{in}(d)$. Recall that $f_{eject}(d)$ in Fig. 3 came from integrating $\Pi_{in}(d)$ according to Eq. (2). For the case without polyvalent counterions and with buffer conditions (0.5M NaCl/10mM TrisCl/1mM EDTA) that differ only slightly from our experimental one, a fit to $\Pi_{in}(d)=f_0\ exp(-d/c)$ with $f_0 = 550,000$ atm and $c = 0.28$nm leads to an analytical expression for $f_{eject}(d)$. Since both $\Pi_{in}(d)$ and $f_{eject}(d)$ are elementary functions of the $d$-spacing between the strands we obtain immediately the force vs pressure plot shown by the dashed curve in Fig. 4. The solid curve shown there is the plot of force $f_{resist}$ vs pressure $\Pi_{out}$ outside the capsid, from Fig. 2b. Recall that partial ejection involves a *force* balance between the packaging stress from inside and the osmotic resistance from outside. It is evident from Fig. 4 that the *pressures* inside and outside are *not* equal.

Consider, for example, the point of force balance for a fully packaged phage without polyvalent salt (marked with the horizontal dotted line at $f = 11$ pN); according to the dashed curve, the inner pressure of the fully loaded capsid is about 77 atm (shown by vertical dotted line 1) while the outer pressure is only of order 10 atm (shown by the vertical dotted line 2 in the same figure). [In our experiment we measured 20 atm outer pressure, $\Pi_{out}$; 20 vs 10, as mentioned above, results from the slightly different solution conditions (lower ionic strength) used in our experiments compared to the ones used to calculate pressure/force here.] The difference (67 atm) in osmotic pressures arises from the fact that the viral capsid is essentially a closed *rigid* object, so that no osmotic pressure equilibrium of the DNA is expected between the inside and the outside even before the addition of an osmotic stress agent. Unlike in the classical osmometer experiment in which an open system allows for volume changes in the sample compartments, thus allowing the spacing between DNA strands to adjust to its equilibrium value, the phage capsid maintains its constant volume where DNA is from the beginning compressed to a *d*-spacing (2.5 nm) below its equilibrium value.

We can also analyze the pressure difference that obtains at the point of force balance for the case where polyvalent ions are present. Using once again the relevant osmotic stress data for $\Pi_{in}(d)$[25b] we plot in Fig. 4 the force versus pressure inside the capsid for the case with 20 mM $Co(NH_3)_6^{3+}$; see dot-dashed curve. Here the force is again calculated by integrating the pressure, but now – unlike the purely repulsive interaction where pressure goes to zero only at infinite spacing –there is a preferred (finite) spacing, $d_0$, where the osmotic pressure vanishes because of the attractive component in the interaction between neighboring portions of DNA. The form of the pressure is now $\Pi_{in}(d) = f_0[\exp((d_o - d)/c) - 1]$ with $f_0$=4.8 atm, $d_0$ = 2.8nm, and $c$= 0.14nm, with the force obtained again from $f_{eject}(d) = \int_d^{d_0} dx \sqrt{3} x \Pi_{in}(x)$. Once again the pressures inside and outside are not balanced. Fig. 4 shows that at the force balance point of 2 pN marked by the lower horizontal dotted line (deduced from Fig. 3 for this set of solution conditions) the pressure inside the viral capsid is about 36 atm (vertical dotted line 3), but the pressure outside is only about 2 atm (vertical dotted line 4). Because of the attractive interaction between the DNA strands introduced by the presence of polyvalent counterion, the pressure difference is about two times smaller [$(\Pi_{in} - \Pi_{out}) = 34$ atm] than in the earlier, simple salt, case [where $(\Pi_{in} - \Pi_{out}) = 67$ atm].

3. CONCLUSION

We have analyzed the forces and pressures inside and outside a bacterial virus, as functions of the length of DNA remaining inside and of the concentrations of salts and osmotic stress polymers in the host solution. Because of the geometry and physical properties of the virus, i.e., a rigid capsid and a long straight tail through which DNA is ejected, the extent of ejection into a colloidal suspension is determined by a balance of inside and outside *forces* rather than pressures. The outside force is determined by the osmotic pressure exerted by the colloidal particles in solution, while the inside force is determined by the confinement stress associated with the crowding of the packed portions of DNA within the capsid. As the ejection proceeds the inside force drops monotonically until $f_{eject}$ has dropped to the value of $f_{resist}$ set by the osmolyte concentration in the external solution. Furthermore, the inside force, $f_{eject}$, for a given length of confined DNA, is set by the ambient values of the simple and polyvalent salt concentrations. For each set of salt conditions one finds a different value of the outside osmotic pressure ($\Pi_{out}^{crit}$) that is sufficient to completely inhibit ejection. In the case of simple salts we find at this force balance point that the pressures inside and outside differ by as much as 70 atm. In the presence of multivalent cations these pressures, and their difference, can be many times smaller.

ACKNOWLEDGEMENTS

We are pleased to acknowledge many helpful discussions with Avinoam Ben-Shaul, Shelly Tzlil, Markus Deserno, Rob Phillips, Prashant Purohit, Mandar Imandar, Bengt Jönsson, Ulf Olsson and Håkan Wennerström. The work has been financially supported by the National Science Foundation, through grants #CHE00-7931 to CMK and #CHE99-88651 to WMG, and by the Swedish Foundation for Internationalization of Research and Higher Education (STINT) through a grant to AE.

REFERENCES:

(1) *Phage and the Origins of Molecular Biology*; Cairns, J.; Stent, G. S.; Watson, J. D., Eds.; Cold Spring Harbor Press, 1992.
(2) Alberts, B.; Bray, D.; Lewis, J.; Raff, M.; Roberts; Watson, J. D. *Molecular Biology of The Cell*, 3 ed.; Garland Publishing: New York, 1994.
(3) Black, L. W. *Ann. Rev. Microbiol.* **1989**, *43*, 267.
(4) Mosing, G.; Eiserling, F. Chapter 9 in *The Bacteriophages*; Calendar, R., Ed.; Plenum: New York, 1988; Vol. 2.
(5) Murialdo, H. *Ann. Rev. Biochem.* **1991**, *60*, 125.
(6) Hershey, A. D.; Chase, M. *J. Gen. Physiol.* **1952**, *36*, 39.
(7) Douglas, T.; Young, M. *Nature* **1998**, *392*, 152.
(8) Slilaty, S. N.; Berns, K. I.; Aposhian, H. V. *J. Biol. Chem.* **1982**, *257*, 6571.
(9) Kindt, J. T.; Tzlil, S.; Ben-Shaul, A.; Gelbart, W. M. *Proc. Nat. Acad. Sci. USA* **2001**, *98*, 13671.
(10) Tzlil, S.; Kindt, J. T.; Gelbart, W. M.; Ben-Shaul, A. *Biophys. J.* **2003**, *84*, 1616.
(11) Purohit, P. K.; Kondev, J.; Phillips, R. *Proc. Nat. Acad. Sci. USA* **2003**, *100*, 3173.
(12) Smith, D. E.; Tans, S. J.; Smith, S. B.; Grimes, S.; Anderson, D. L.; Bustamante, C. *Nature* **2001**, *413*, 748.
(13) Evilevitch, A.; Lavelle, L.; Knobler, C. M.; Raspaud, E.; Gelbart, W. M. *Proc. Nat. Acad. Sci. USA* **2003**, *100*, 9292.
(14) Serwer, P. *Biopolymers* **1988**, *27*, 165.
(15) Mitchell, P.; Moyle, J. *Symp. Soc. Gen. Microbiol.* **1956**, *6*, 150.
(16) Stock, J. B.; Rauch, B.; Roseman, S. *J. Biol. Chem.* **1977**, *252*, 7850.
(17) Roa, M. *FEMS Microbiology Letters* **1981**, *11*, 257.
(18) Graff, A.; Sauer, M.; Van Gelder, P.; Meier, W. *Proceedings of the National Academy of Sciences of the United States of America* **2002**, *99*, 5064.
(19) Novick, S. L.; Baldeschwieler, J. D. *Biochemistry* **1988**, *27*, 7919.
(20) Evilevitch, A.; Gober, J. W.; Phillips, M.; Knobler, C. M.; Gelbart, W. M. *manuscript in preparation* **2004**.
(21) Lau, A. W. C.; Pincus, P. A. *P. Phys. Rev.* **2002**, *E66*, 041501.
(22) Nguyen, T. T.; Rouzina, I.; Shklovskii, B. I. *J. Chem. Phys.* **2000**, *112*,

2562.

(23) Solis, F. J.; Olvera de la Cruz, M. *Phys. Rev.* **1999**, *E60*, 4496.

(24) Arenzon, J. J.; Stilck, J. F.; Levin, Y. *Eur. Phys. J.* **1999**, *B12*, 79.

(25) a. Parsegian, V. A.; Rand, R. P.; Fuller, N. L.; Rau, D. C. *Methods in Enzymology* **1986**, *127*, 400. Rau, D. C.; b. Parsegian, V. A. *Biophys. J.* **1992**, *61*, 246. Leikin, S.; c. Parsegian, V. A.; Rau, D. C. *Annu. Rev. Phys. Chem.* **1993**, *44*, 369.

(26) a. Raspaud, E.; de la Cruz, M. O.; Sikorav, J. L.; Livolant, F. *Biophys. J.* **1998**, *74*, 381. b. Pelta, J.; Livolant, F.; Sikorav, J.-L. *J. Biol. Chem.* **1996**, *271*, 5656.

(27) Wilson, R. W.; Bloomfield, V. A. *Biochemistry* **1979**, *18*, 2192.

(28) Baldwin, R. L.; Widom, J. *Biopolymers* **1983**, *22*, 1595.

(29) Evilevitch, A.; Grayson, P.; Phillips, R.; Knobler, C. M.; Gelbart, W. M. *manuscript in preparation* **2004**.

(30) Castelnovo, M.; Bowles, R. K.; Reiss, H.; Gelbart, W. M. *Eur. Phys. J. E.* **2003**, *10*, 191.

(31) Maniatis, T.; Fritsch, E. F.; Sambrook, J. *Molecular Cloning A Laboratory Manual*, seventh ed.; Cold Spring Harbor Laboratory, 1983.

(32) de Gennes, P. G. *Scaling Concenpts in Polymer Physics*; Corenll University: Ithaca: N. Y., 1979. See also: de Gennes, P.G. *C. R. Acad. Sci. B.* **1979**, *288*, 359.

(33) de Vries, R. *Biophys. J.* **2001**, *80*, 1186.

(34) Abbott, N. L.; Blankschtein, D.; Hatton, T. A. *Macromolecules* **1992**, *25*, 3917.

(35) Earnshaw, W. C.; Casjens, S. R. *Cell* **1980**, *21*, 319.

(36) Earnshaw, W. C.; Harrison, S. C. *Nature* **1977**, *268*, 598.

FIGURE CAPTIONS:

**Figure 1**
Schematic of the experiment in which viral DNA ejection is inhibited by the osmotic force due to PEG in the external solution. The force balance is measured at equilibrium after all the ejected DNA has been digested by DNaseI. The ejected DNA is pictured as a short stiff chain that extends into the host solution. For the sake of clarity, the proteins LamB and DNase I have not been pictured.

**Figure 2**
Analysis of the experimental data to deduce the ejection force for different lengths of external DNA. (a) Experimental results: percentage of ejected DNA as a function of $\Pi_{out}$, the osmotic pressure (atm) of the surrounding PEG solution in the spermine-free (circles and solid curve) and spermine-added (diamonds and dot-dashed curve) cases. In each case only one set of data[13] (from one batch of phage) has been shown and error bars have been omitted for clarity. (b) Theoretical results for $f_{resist}$, the force (pN) resisting DNA insertion into the PEG solution as a function of the $\Pi_{out}$, the external osmotic pressure (atm), for degree of polymerization $M=130$ and monomer size $b=0.4nm$. (c) Combination of the two previous plots to give the ejection force (under equilibrium conditions) as a function of the DNA fraction ejected for both cases: with (dot-dashed curve) and without spermine (solid curve).

**Figure 3**
Ejection force (pN) as a function of interaxial spacing (nm), as calculated from Eq. (2), for two different sets of solution conditions, assuming hexagonal packing of DNA. The first set of conditions (dashed curve), 0.5M NaCl/10mM TrisCl/1mM EDTA, mimics the ejection experiment in the absence of spermine. The second set (dot-dashed curve), 20mM Co(NH$_3$)$_6$Cl$_3$/0.25M NaCl/10mM TrisCl, shows the effect of the addition of polyvalent ions, with composition close to that of our spermine-added experiment. The dotted horizontal lines show the forces in the two cases when the interaxial distance is 2.49 nm, which corresponds to the measured values in the lambda mutant used here (86% of wild-type lambda genome).

**Figure 4**
Force (pN) as a function of the outside pressure ($\Pi_{out}$: top, solid curve) and the inside pressure ($\Pi_{in}$: bottom two). The force outside, $f_{resist}$, is calculated as described for Fig. 2b. The force inside, $f_{eject}$, is calculated by assuming hexagonal packing of DNA using the Rau-Parsegian measurements for the two solution conditions discussed in Fig. 3 (dashed and dot-dashed curves). The vertical dotted lines (1) and (2), and (3) and (4), show the pressures inside and outside at maximum ejection force without (see lines 1 and 2), and with (3 and 4), added polycation. Note the large differences in pressures (67 atm and 34 atm, respectively) when the system is at force equilibrium.

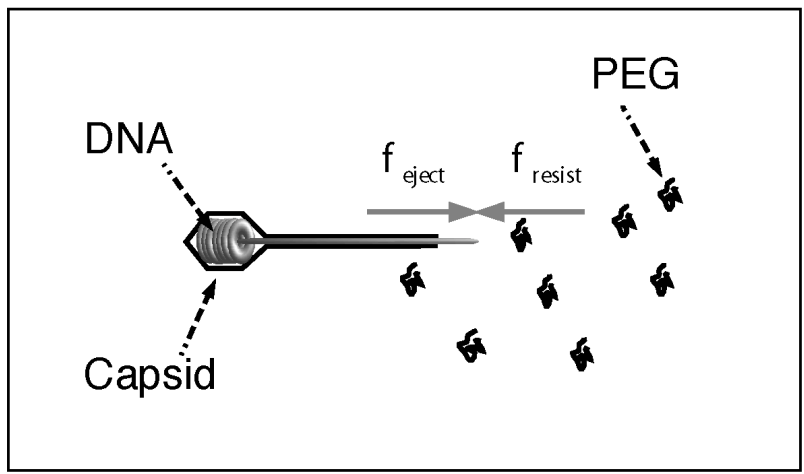

Figure 1

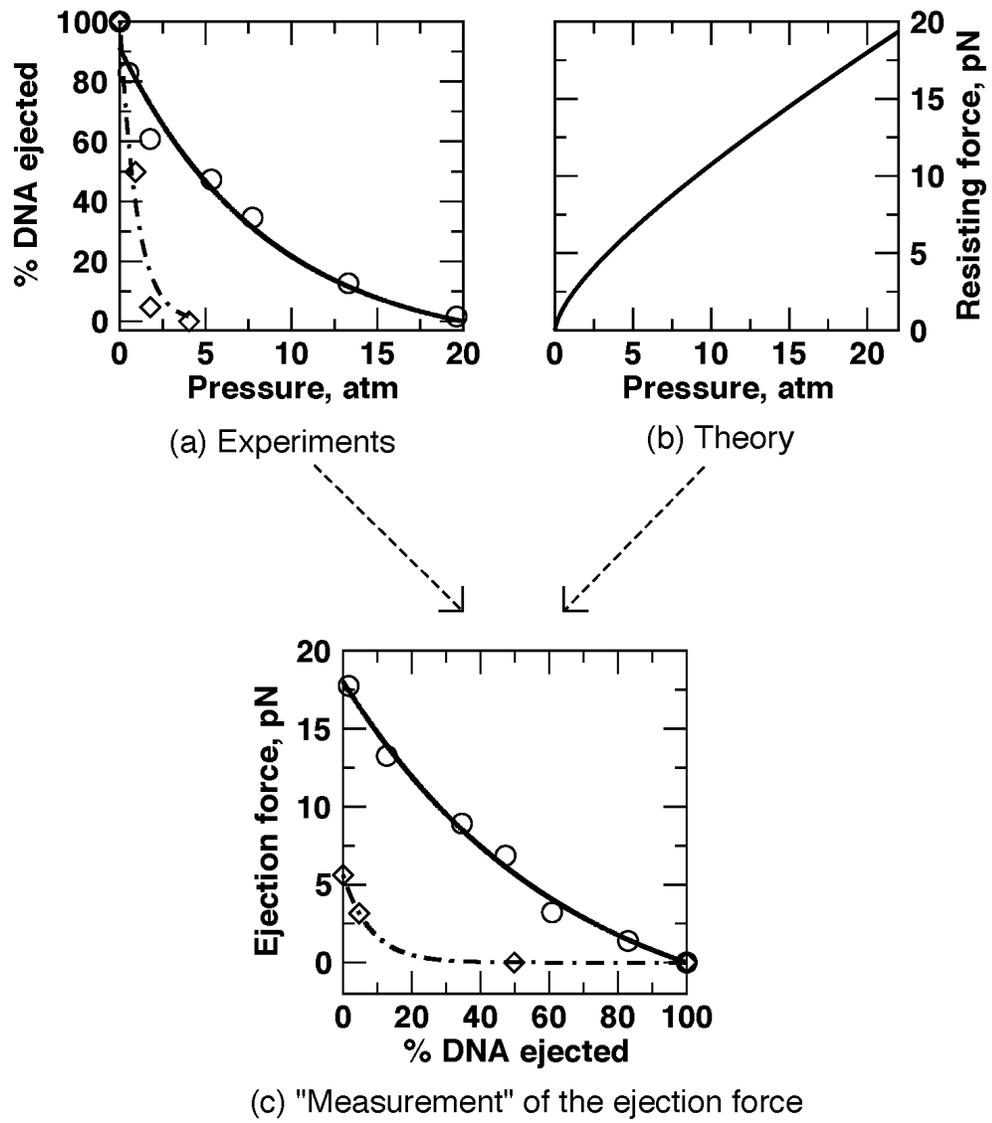

(c) "Measurement" of the ejection force

Figure 2

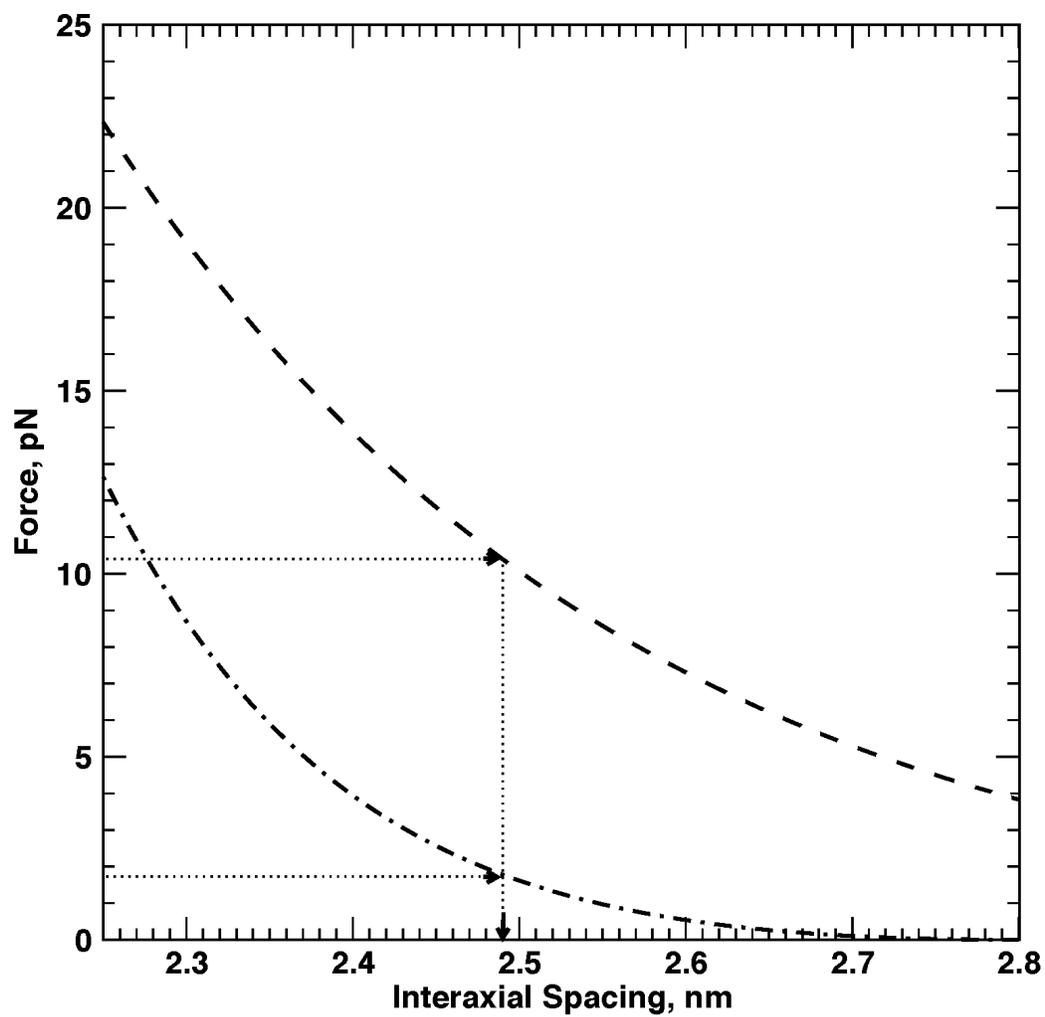

Figure 3

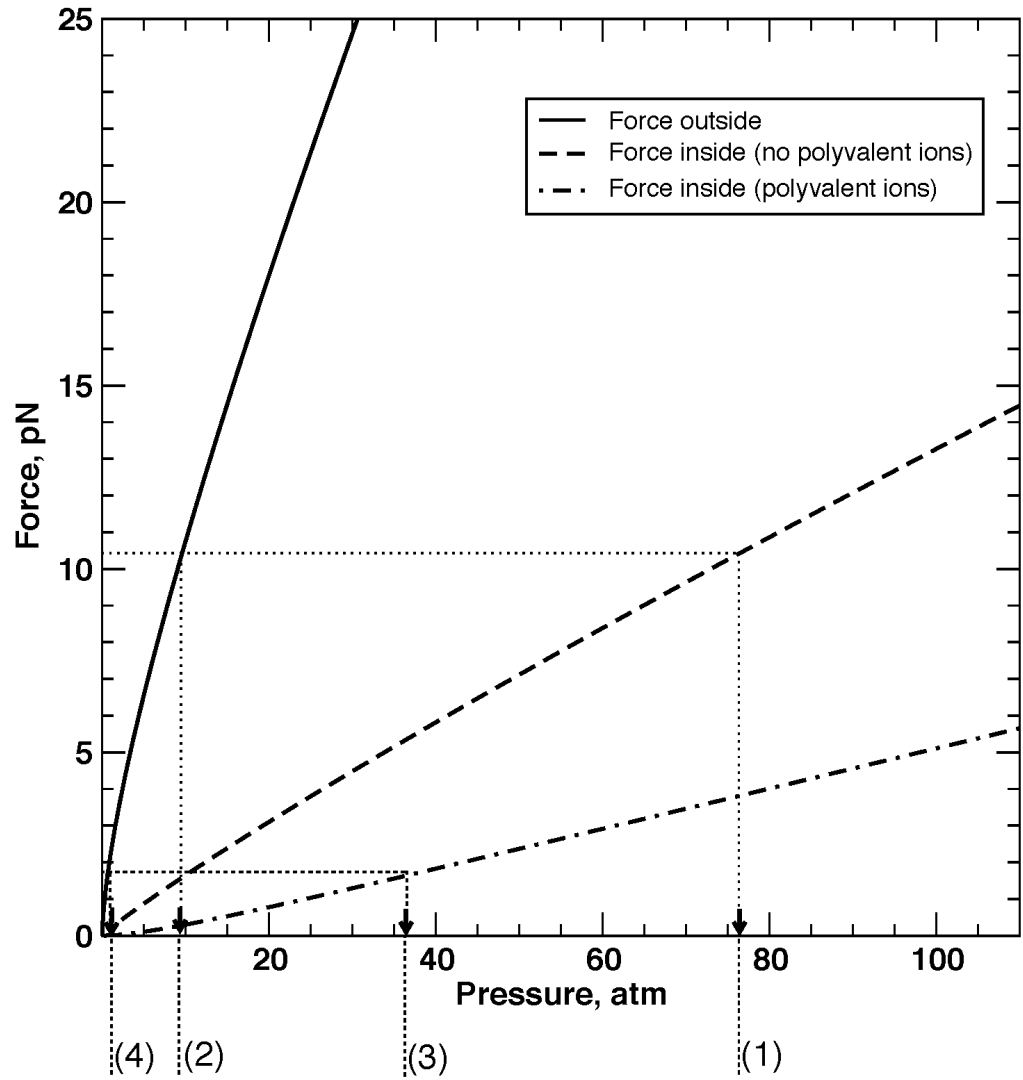

Figure 4